\documentclass[11pt,a4paper,superscriptaddress,prb,amsmath,amssymb,aps]{revtex4}
\usepackage{graphicx,graphicx,epsfig} 

\usepackage{graphicx}
\usepackage{bm}

\begin{document}
\title{Geometric phase for non-Hermitian Hamiltonian evolution as anholonomy of a parallel transport along a curve. } 

\author{N.A. Sinitsyn}
\affiliation{Center for Nonlinear Studies and Computer, Computational and Statistical Sciences Division, Los Alamos National Laboratory, Los Alamos, NM 87545 USA}

\author{Avadh Saxena}
\affiliation{Theoretical  Division, Los Alamos National Laboratory, Los Alamos, NM 87545 USA}


\begin{abstract}
We develop a new interpretation of the geometric phase in evolution with a non-Hermitian real value Hamiltonian 
by relating it to the angle developed during the parallel transport along a closed curve by a unit vector triad in the 3D-Minkovsky space.
We also show that this geometric phase is responsible for the anholonomy effects in stochastic processes
considered in [N. A. Sinitsyn and I. Nemenman, EPL {\bf 77}, 58001 (2007)], and use it to derive the stochastic system 
response to periodic parameter variations.
\end{abstract}

\date{\today}

\maketitle 

In quantum mechanics, anholonomy effects (i.e. parallel transported vectors not returning 
to their initial orientations after a motion along a closed curve), usually can be related to the Berry phase \cite{berry-84}.
Similar effects have been recognized in many other fields and  were also related to several generally defined geometric phases. Examples can be found in
classical mechanics \cite{hannay-85,littlejohn-88}, hydrodynamics \cite{wilczek-88}, classical chaos \cite{jarzynski-chaos}, soliton dynamics \cite{marsden-soliton}, 
dissipative kinetics \cite{kagan-91, landsberg-92,sinitsyn-08jpa}, and stochastic processes \cite{sinitsyn-07epl, sinitsyn-07prl, sinitsyn-07prb, ohkubo-08, shi, astumian-pnas}.

Simple systems, with a minimal number of degrees of freedom have always been of particular importance. Thus the essential features of 
the Berry phase in quantum mechanics can be discussed using a two-level system and the corresponding $SU(2)$ group of its  
evolution.  
Another simple group of transformations, which was widely discussed in relation to geometric phases, is the SU(1,1) group, and isomorphic to it SL(2,R). It is also  homomorphic to the
Lorentz group SO(2,1) \cite{lie-group-book}. The essential difference with respect to $SU(2)$ case is that the quotient manifold $SU(1,1)/U(1)$ can be identified with a
hyperboloid rather than a sphere.  Corresponding geometric phases have been predicted and studied in several  classical mechanical, relativistic, and quantum mechanical applications
 \cite{hannay-85,sl2r-1,sl2r-2,sl2r-3,sl2r-4,mukunda-03jpa,ferraro-00}, and were also measured in experiments on polarized light propagation \cite{sl2r-optics1,sl2r-optics2,sl2r-optics3}. 

The SU(2) Berry phase anholonomy
can be nicely explained by relating it to the rotation angle of a unit vector triad, associated with a closed curve drawn by a unit Bloch vector
on a sphere \cite{urbantke,dandoloff-89jpa,dandoloff-92jpa} (for a textbook demonstration see also \cite{berry-book}).
Similar formulation was proved to be useful in other contexts,
e.g. in the motion of charged particles in a nonuniform magnetic field \cite{littlejohn-88}, light propagation \cite{berry-nature}, or a motion in a noninertial frame \cite{berry-book}.
In Ref.~\onlinecite{dandoloff-92jpa} it was employed to derive new inequalities for the evolution with the SU(2) group.
However, to our knowledge, similar interpretation of the non-Hermitian SL(2,R) evolution has not been explicitly presented, although it is expected, cosidering the well known relation of the SL(2,R) group 
and the Lorenz group.

In this communication we show exactly how the SL(2,R) geometric phase can be illustrated as the anholonomy of the parallel trasport of a vector frame,
 with a vector triad, defined in the 3D Minkovsky space with correspondingly defined 
vector multiplication rules. An additional goal
is to show that the recently introduced geometric phases in purely classical stochastic kinetics \cite{sinitsyn-07epl,sinitsyn-07prb} provide one more application of the SL(2,R) geometric phase. 
We will use this fact to determine the geometric contribution to particle currents in a model proposed in Ref.~\onlinecite{sinitsyn-07epl}, assuming time dependence of all parameters.

Consider the evolution of a real two state vector $\vert u \rangle = (u_1,u_2)$ according to the equation
\begin{equation}
\frac{d}{dt} \vert u(t) \rangle 
= \hat{H}(t) 
\vert u(t) \rangle, \quad H(t) = \left( \begin{array}{ll} 
h_{11}(t)  &  h_{12}(t)\\
h_{21}(t)    & h_{22}(t)
\end{array} \right),
\label{slev}
\end{equation}
with slowly time-dependent real parameters $h_{ij}(t)$, $i,j=1,2$. Formally, solution of (\ref{slev}) can be written as a time ordered exponent of the time-integral of $\hat{H}(t)$
  \begin{equation}
  \vert u(t) \rangle  = \hat{U} \vert u(0) \rangle, \quad \hat{U}= \hat{T} \left[ e^{\int_0^t \hat{H}(t)dt} \right]. 
\label{slev-sol}
\end{equation}
If the matrix $\hat{H}$ were traceless ($h_{11}=-h_{22}$) the evolution matrix $\hat{U}$ would belong to the group $SL(2,R)$, i.e. the class of 2$\times$2 matrices with real entries and a unit determinant. The requirement 
to have a zero trace of $\hat{H}$, however, is not crucial for the following discussion, because the nonzero trace merely shifts the eigenvalues of this matrix but does not change its eigenvectors. Therefore
the geometric phase is not sensitive to this property, so we will refer to the geometric phase of the group SL(2,R)  even if $\hat{H}$ has a nonzero trace.  

For the future discussion we will also consider  the left state vector $\langle v \vert  =(v_1,v_2)$, evolving according to 
\begin{equation}
\frac{d}{dt}\langle v\vert =-\langle v \vert \hat{H}, \quad \langle v(t) \vert u(t) \rangle=1.
\label{conj}
\end{equation} 
Let the matrix $\hat{H}$ have two real eigenvalues. For adiabatically slow evolution of parameters, the right eigenvector corresponding to the 
larger eigenvalue $\lambda_0$ will completely dominate over the other one. If the evolution starts from this eigenvector and parameters pass through a cycle the final vector
will return to the initial one, however it will be multiplied by an overall factor $e^{\phi}$, i.e.
 \begin{equation}
  \left( \begin{array}{l}
u_1(T)\\
u_2(T)
\end{array} \right) = e^{\phi} 
\left( \begin{array}{l}
u_1(0)\\
u_2(0)
\end{array} \right). 
\label{slev-sol}
\end{equation}
The ``phase" $\phi$ is not imaginary, however, a lot of analogies with quantum mechanical Berry phase can be established. The Berry phase was generalized to a 
non-Hermitian evolution \cite{nonherm-bf1,berry-nonherm}, and the well established result is that in the 
adiabatic limit the phase $\phi$  can still be written as a sum of dynamical and geometric contributions, i.e.
$\phi=\phi_d+\phi_g$, where $\phi_d = \int_0^T \lambda_0(t) dt$.
The expression for the geometric phase can be written as a parallel transport condition. For this 
 one should redefine states $ |u(t) \rangle  \rightarrow  e^{-\int_0^t\lambda_0(t)dt} |u (t) \rangle$, and 
$ \langle v(t) | \rightarrow  e^{\int_0^t\lambda_0(t)dt} \langle v (t)|$. 
 The geometric phase $\phi_g$  can then be expressed as arising from the condition \cite{urbantke}
\begin{equation}
\langle v(t)| \partial_t u(t) \rangle =0,
\label{par1}
\end{equation}
i.e. if we assume that $\vert u(t) \rangle = e^{\phi_g(t)} \vert u(\{h_{ij} \} \rangle$, and $\langle v(t) \vert = e^{-\phi_g(t)} \langle v(\{h_{ij} \} \vert$, where $\vert u(\{h_{ij} \} \rangle$, and 
$ \langle v(\{h_{ij} \} \vert$ are instantaneous gauge fixed normalized right and left eigenstates, corresponding to the same eigenvalue $\lambda_0 (\{ h_{ij} \})$ of the matrix $\hat{H}$,
then the geometric phase after completion of the cyclic evolution reads     
\begin{equation}
\phi_g=-\oint dt \langle v(\{ h_{ij} \}) \vert \partial_t u(\{ h_{ij} \}) \rangle.
\label{phi_g}
\end{equation}
Urbantke \cite{urbantke} showed that for a quantum mechanical spin-1/2 the condition analogous to (\ref{par1}) has a simple geometrical interpretation in terms
of a parallel transport of a unit vector triad. Now we show that a similar interpretation is possible for the SL(2,R) group, however, the triad should be
defined in the 3D Minkovsky space.   
 
\begin{figure}
\includegraphics[width=7.5 cm]{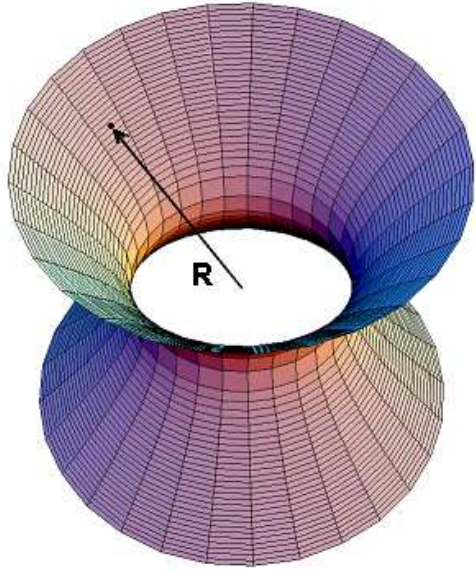}
\centering
\caption{\label{hyperboloid} Hyperboloid representing possible states of the vector ${\bf R}$. }
\end{figure}

Components of the right and left vectors $\vert  u \rangle$ and $\langle v \vert$ can be used to compose a vector ${\bf R}$ such as
\begin{equation}
{\bf R}=(x,y,z)=(u_1v_1-u_2v_2, v_2u_1+v_1u_2, v_2u_1-v_1u_2).
\label{rad}
\end{equation}
The normalization condition in (\ref{conj}) then leads to the following normalization of ${\bf R}$
\begin{equation}
{\bf R}\,\,\widetilde{\cdot} \,\,{\bf R}=x^2+y^2-z^2=1,
\label{norm3}
\end{equation}
where we introduced a scalar product operation in the 3D Minkovsky space ${\bf a}\,\,\widetilde{\cdot} \,\,{\bf b} \equiv a_1b_1+a_2b_2 -a_3b_3$.
Fig.~\ref{hyperboloid} shows that vector ${\bf R}$ can be represented by a point on a unit hyperboloid immersed in the 3D Minkovsky space.
%
%
Lets introduce
\begin{equation}
{\bf P}= (-2u_1u_2, u_1^2-u_2^2, u_1^2+u_2^2),\quad {\bf Q}=(-2v_1v_2,v_1^2-v_2^2,-(v_1^2+v_2^2)),
\label{pq}
\end{equation}
and compose two more vectors out of them
\begin{equation}
{\bf N}= ({\bf P}+{\bf Q})/2,\quad {\bf S}=({\bf P}-{\bf Q})/2.
\label{ns}
\end{equation}
One can  check that  ${\bf R}$, ${\bf N}$ and ${\bf Q}$ are mutually orthogonal with respect to the metric $(+,+,-)$, namely
\begin{equation}
\begin{array}{l}
{\bf R} \,\widetilde{\cdot} \,\, {\bf S}={\bf R} \,\,\widetilde{\cdot} \,\, {\bf N}={\bf S} \,\,\widetilde{\cdot} \,\, {\bf N}=0,\\
{\bf R} \,\,\widetilde{\cdot} \,\, {\bf R}={\bf N} \,\,\widetilde{\cdot} \,\, {\bf N}=1,\\
{\bf S} \,\,\widetilde{\cdot} \,\, {\bf S}=-1.
\end{array}
\label{triad1}
\end{equation}
Vectors ${\bf R}, {\bf N}$, and ${\bf S}$ comprise a unit triad in the 3D Minkovsky space, such that vectors ${\bf R}$ and ${\bf N}$ are space-like and 
${\bf S}$ is time-like. 

What does the parallel transport condition (\ref{par1}) mean for the evolution of the triad? For the components of $\vert u \rangle$ and $\langle v \vert$ it means that 
$v_1\partial_tu_1+v_2\partial_tu_2=0$. Following Urbantke \cite{urbantke} this suggests that $\partial_tu_1 =-\lambda_1 v_2$, $\partial_t v_1=-\lambda_2 u_2$, $\partial_t u_2 = \lambda_1 v_1$, and
$\partial_t v_2 = \lambda_2 u_1$, with some variables $\lambda_1$ and $\lambda_2$ that depend on the details of the evolution Hamiltonian. Substituting this into the definition of the
triad vectors we find that
  \begin{equation}
  \frac{d}{dt}\left( \begin{array}{l}
{\bf N}\\
{\bf R} \\
{\bf S}
\end{array} \right) = \left( \begin{array}{lll}
\,\,\, 0 & \tau & 0 \\
-\tau & 0 & \varkappa \\
\,\,\, 0 & \varkappa & 0
\end{array} \right)
\left( \begin{array}{l}
{\bf N}\\
{\bf R}\\
{\bf S}
\end{array} \right), \quad \tau=-(\lambda_1+\lambda_2),\quad \varkappa=\lambda_2-\lambda_1.
\label{triad-ev}
\end{equation}
Conditions (\ref{triad-ev}) have the form of Serret-Frenet equations in 3D Minkovsky spacetime. According to Ref.~\onlinecite{minkovsky-tetrad} they describe a
 unique regular curve  parametrized by $t$, with a curvature $\varkappa$ and torsion
 $\tau$. From
$\dot{{\bf N}}=\tau {\bf R}$ and ${\bf N}\cdot {\bf R}=0$, it follows that $\tau=\dot{{\bf N}}\,\,\widetilde{\cdot} \,\,{\bf R} = -\dot{{\bf R}} \,\,\widetilde{\cdot} \,\, {\bf N}$, and a similar relation holds for $\varkappa$ in terms
of ${\bf S}$ and $\dot{{\bf R}}$, which substituted back in (\ref{triad-ev}) results in 
\begin{equation}
\dot{\bf N}=-({\bf N} \,\,\widetilde{\cdot} \,\, \dot{{\bf R}}){\bf R},\quad \dot{\bf S}=-({\bf S} \,\,\widetilde{\cdot} \,\, \dot{{\bf R}}){\bf R}.
\label{triad-ev2}
\end{equation}

 This type of vector evolution is a special case of the Fermi-Walker vector transport in special relativity. It 
can be interpreted as follows. Suppose, vectors ${\bf N}$ and ${\bf S}$ at point ${\bf R}(t+dt)$ are obtained by translation of
vectors ${\bf N}({\bf R}(t))$ and ${\bf S}({\bf R}(t))$ to the point ${\bf R}(t+dt)$ that is followed by a projection
onto the 2D subspace of vectors orthogonal to ${\bf R}(t+dt)$. Up to higher order in $dt$, one can write ${\bf N}(t) \,\,\widetilde{\cdot} \,\,
 {\bf R}(t+dt) \approx {\bf N} \,\,\widetilde{\cdot} \,\, \dot{{\bf R}}(t) dt$, and a similar relation holds for ${\bf S}$. 
Then ${\bf N}(t+dt)={\bf N}(t)-{\bf N}(t)\,\,\widetilde{\cdot} \,\, \dot{{\bf R}}(t) dt$, and ${\bf S}(t+dt)={\bf S}(t)-{\bf S}(t)\,\,\widetilde{\cdot} \,\, \dot{{\bf R}}(t) dt$.  
This means that the conditions (\ref{par1}) and (\ref{triad-ev2}) correspond to the parallel transport of vectors ${\bf N}$ and ${\bf S}$ along the curve ${\bf R}(t)$. 

Parallel transported vectors generally do not return to the initial ones after a motion along a closed curve, which represents the anholonomy effect. The relation of such an anholonomy to the geometric phase
can be inferred if we observe how these vectors change under the gauge transformations $\vert u' \rangle = e^{\phi} \vert u \rangle$ and $\langle v' \vert = \langle v \vert e^{-\phi}$. This 
corresponds to ${\bf P'} = {\bf P}e^{2\phi}$ and ${\bf Q'} = {\bf Q}e^{-2\phi}$. The triad transformation then reads
\begin{equation}
\begin{array}{l}
{\bf R'} = {\bf R}, \\
{\bf N'} = {\bf N} \cosh (2\phi) + {\bf S} \sinh (2\phi), \\
{\bf S'} = {\bf N} \sinh (2\phi) + {\bf S} \cosh (2\phi),
\end{array}
\label{boost}
\end{equation}
which indicates that the vector ${\bf R}$ is gauge invariant, but the vectors ${\bf N}$ and ${\bf S}$ are mixed with each other like after a boost transformation
in the Minkovsky space. The normalization properties (\ref{triad1}), however, remain unaltered. This result means that if after the parallel transport along a closed curve
the vectors ${\bf N}$ and ${\bf S}$ become mixed with the angle $\phi$, it corresponds to a multiplication of the state vector $\vert u \rangle$ by an exponential geometric phase
factor $exp(\phi_g)$, where
\begin{equation}
\phi_g=\phi/2.
\label{geomphi}
\end{equation}

To derive the geometric phase, it is thus sufficient to compare the rotation of the parallel transported vectors ${\bf N}$ and ${\bf S}$ to a pair of fixed reference vectors.
Let us introduce a vector product operation $({\bf a} \,\,\widetilde{\times} \,\, {\bf b})_i=g_{ik}\epsilon^{ksm}a_sb_m$, where $g_{ik}$ is the metric tensor of the 3D Minkovsky space with signature $(+,+,-)$, and 
$\epsilon^{ksm}$ is the Levy-Civita symbol. 
 It
is possible to assign the fixed triad field ${\bf e_1}$, ${\bf e_2}$, and ${\bf e_3}$ in the Minkovsky space as follows.
\begin{equation}
{\bf e_3}=(0,0,1),\quad {\bf e_1}=\frac{{\bf R} \,\,\widetilde{\times} \,\, {\bf e_3}} {|{\bf R} \,\,\widetilde{\times} \,\, {\bf e_3}|},\quad {\bf e_2}=\frac{{\bf R} \,\,\widetilde{\times} \,\, {\bf e_1}}{R},
\label{fixedtriad}
\end{equation} 
 where $R=\sqrt{x^2+y^2-z^2}$. Explicitly,
 \begin{equation}
 {\bf e_1}=\left( \frac{y}{\sqrt{x^2+y^2}},\frac{-x}{\sqrt{x^2+y^2}},0 \right),\quad
 {\bf e_2}=\frac{1}{R}\left( \frac{xz}{\sqrt{x^2+y^2}}, \frac{yz}{\sqrt{x^2+y^2}} , \sqrt{x^2+y^2} \right).
\label{fixedtriad2}
\end{equation} 
It is straightforward to show that
\begin{equation}
\begin{array}{l}
{\bf R} \,\,\widetilde{\cdot} \,\, {\bf e_1}={\bf R} \,\,\widetilde{\cdot} \,\, {\bf e_2}={\bf e_1} \,\,\widetilde{\cdot} \,\,{\bf e_2}=0\\
{\bf e_1} \,\,\widetilde{\cdot} \,\, {\bf e_1}=-{\bf e_2} \,\,\widetilde{\cdot} \,\, {\bf e_2}=1
\end{array}
\end{equation}
Consequently, ${\bf e_1}$ and ${\bf e_2}$ provide a pair of orthogonal unit vectors in the space orthogonal to ${\bf R}$. Vector ${\bf e_1}$ is space-like and
${\bf e_2}$ is time-like. Note that although corresponding vector fields are fixed, the local frame ${\bf e_1}({\bf R}(t))$ and ${\bf e_2}({\bf R}(t))$ will
depend on $t$ for an  observer, moving along a trajectory ${\bf R}(t)$.
During the parallel transport the pair ${\bf N, S}$ would also rotate around ${\bf e_1,e_2}$,
  \begin{equation}
\left( \begin{array}{l}
{\bf N}(t)\\
{\bf S}(t)
\end{array} \right) = \left( \begin{array}{lll}
\cosh (\phi(t)) & \sinh(\phi(t)) \\
\sinh(\phi(t))  & \cosh(\phi(t))
\end{array} \right)
\left( \begin{array}{l}
{\bf e_1}({\bf R}(t))\\
{\bf e_2}({\bf R}(t))
\end{array} \right) .
\label{triad-ev-3}
\end{equation}
From the parallel transport conditions, it follows that 
\begin{equation}
{\bf N} \,\,\widetilde{\cdot} \,\, d{\bf S}=0.
\label{part3}
\end{equation}
Substituting (\ref{triad-ev-3}) into (\ref{part3}) and then using (\ref{fixedtriad2}) we find that this leads to
\begin{equation}
2d\phi_g=d \phi = -{\bf e_1} \,\,\widetilde{\cdot} \,\, d{\bf e_2}=-\frac{zydx-zxdy}{R(x^2+y^2)}.
\label{dphi}
\end{equation}
The geometric phase acquired after the motion of vector ${\bf R}$ along a closed contour  can then be written as
\begin{equation}
\phi_g=\oint_{\bf c} {\bf A}\cdot d{\bf R} =\iint_{S_{\bf c}} F,
\label{geomphase}
\end{equation} 
where ${\bf A}=(-\frac{zy}{2R(x^2+y^2)},\frac{zx}{2R(x^2+y^2)},0)$,
and in the last step we used the Stokes theorem to  express a contour integral along ${\bf c}$ as an integral  over the surface $S_{\bf c}$ inside this contour from the Berry curvature. The latter,
on the surface of the unit hyperboloid ($R=1$), explicitly reads
\begin{equation}
F=-\frac{1}{2} \left( x dy\wedge dz+y  dz \wedge dx +z dx\wedge dy \right).
\label{berrycurvature}
\end{equation}
This curvature 2-form is well known in relation to the groups $SU(1,1)$ and $SL(2,R)$ \cite{sl2r-3}. Our derivation, however, presents a simple illustration
of the geometric origin of this Berry curvature.

To  switch from the integration over the surface inside ${\bf R}(t)$ to the integral over the surface in the parameter space $\{ h_{ij} \}$, note that
the vector ${\bf R}$ satisfies the Bloch equation \cite{sl2r-optics1} 
\begin{equation}
\frac{d{\bf R}}{dt}={\bf \xi}\,\, \widetilde{\times} \,\, {\bf R}, 
\label{rxi}
\end{equation}
where 
\begin{equation}
{\bf \xi}(\{ h_{ij} \})=-(h_{11}-h_{22},h_{12}+h_{21},h_{12}-h_{21}).
\label{components1}
\end{equation}
The quasi-steady state solution corresponds to 
\begin{equation}
{\bf R}(\{ h_{ij} \})=-{\bf \xi}(\{ h_{ij} \})/|{\bf \xi} (\{ h_{ij} \})|. 
\label{components2}
\end{equation}

\begin{figure}
\includegraphics[width=8.5 cm]{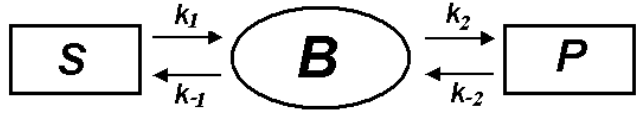}
\centering
\caption{\label{system} Transition rates into and out of the absorbing
  states $S$ (substrate) and $P$ (product) through an intermediate bin $B$-system. The bin can have only zero or one particle inside it.}
\end{figure}

As an example of a new application of SL(2,R) formalism, we consider the geometric 
 phase that was found in a purely classical stochastic system.
 Authors of Ref.~\onlinecite{sinitsyn-07epl} analyzed stochastic particle fluxes from Right to Left reservoirs through an intermediate bin-system with exclusion 
interactions, i.e. allowing at most one particle to be inside the bin. Kinetic rates are shown in Fig.~\ref{system}. The moments generating function of the particle current is defined as \cite{comment}
\begin{equation}
Z(\chi,t)=e^{S(\chi,t)}=
\sum_{n=-\infty}^{\infty} P_{n}e^{n\chi}.
\label{pgf1}
\end{equation}
where $P_{n}$ is the probability to find a total of $n$ particles transfered from Left to Right during the observation time $t$.
Authors of Ref.~\onlinecite{sinitsyn-07epl} showed that (\ref{pgf1}) can be expressed as the average of the evolution operator
\begin{equation}
Z(\chi,t)= {\bf 1}^+\hat{T}\left(e^{-\int_{0}^{t}\hat{H}(\chi,t) dt}\right) {\bf p}(0),
\label{pdf2}
\end{equation}
where
\begin{equation}
\hat{H}(\chi,t)=\left(
\begin{array}{cc}
k_1 + k_{-2}   & -k_{-1}-k_2 e^{\chi} \\
-k_1-k_{-2}e^{-\chi}   & k_{-1}+k_2
\end{array} \right),
\label{hchi}
\end{equation}
${\bf 1}^+=(1,1)$, and ${\bf p}(0)=(p_1,p_2)$ is the vector of initial probabilities of the bin states. 
Up to a matrix proportional to the unit one, the matrix $\hat{H}(\chi,t)$ in (\ref{hchi}) belongs to a set of 
generators of the SL(2,R) group, thus allowing us to apply all known results for this group to expression (\ref{pdf2}).
  
Suppose, parameters $k_1$ and $k_{-2}$ evolve around a closed contour. From the above discussion it follows that after completing the cycle, the 
moments generating function becomes an exponent of the sum of two terms  
\begin{equation}
Z(\chi)=e^{S_{geom}(\chi)+S_{qst}(\chi)},
\label{mgf3}
\end{equation}
where, $S_{qst}(\chi)$ is the quasistationary cumulants generating function averaged over all parameter values along the contour, and $S_{geom}$ is the geometric phase contribution responsible
for additional pump currents. It can be written as an integral over the surface inside the contour created by the curve in the parameter space. 

\begin{equation}
S_{geom}(\chi)=\iint_{S_{\bf c}} F_{k_1,k_{-2}} dk_1dk_{-2}.
\label{bercur}
\end{equation}

Having the general result for the SL(2,R) group (\ref{geomphase}), it is now straightforward to find the Berry curvature in (\ref{bercur}) by a simple change of variables, i.e. 
\begin{equation}
F_{k_1,k_{-2}}({\bf k})=-\frac{1}{2} \left[ x({\bf k})\frac{\partial(y,z)}{\partial (k_1,k_{-2})}+y({\bf k})\frac{\partial(z,x)}{\partial (k_1,k_{-2})}+z({\bf k})\frac{\partial(x,y)}{\partial (k_1,k_{-2})} \right]=
\frac{e_{-\chi}(e^{\chi}k_2+k_{-1})}{[4\kappa_+e_\chi+
4\kappa_-e_{-\chi}+K^2]^{3/2}},
\label{bercur2}
\end{equation}
where components of ${\bf R}$ were taken from (\ref{components1}) and (\ref{components2}),  $\kappa_{\pm} \equiv k_{\pm1}k_{\pm2}$, $e_{\pm\chi} \equiv e^{\pm \chi}-1$, $ K\equiv \sum_{m}k_m$. The Berry curvature in
 (\ref{bercur2}) is the same as the one derived in Ref.~\onlinecite{sinitsyn-07epl}.
%
%
It is  now easy to derive other previously unknown components of the Berry curvature tensor by a similar change of variables
\begin{equation}
\begin{array}{l}
F_{k_1,k_{2}}({\bf k})=
-\frac{e_{\chi}(k_{-1}-k_{-2})}{[4\kappa_+e_\chi+
4\kappa_-e_{-\chi}+K^2]^{3/2}},
\quad F_{k_2,k_{-2}}({\bf k})=-\frac{e_{-\chi}(e^{\chi}k_1+k_{-1})}{[4\kappa_+e_\chi+
4\kappa_-e_{-\chi}+K^2]^{3/2}},\\
\\
 F_{k_{1},k_{-1}}({\bf k})=-\frac{e_{-\chi}(e^{\chi}k_2+k_{-2})}{[4\kappa_+e_\chi+
4\kappa_-e_{-\chi}+K^2]^{3/2}}, \quad F_{k_{-1},k_{-2}}({\bf k})=
\frac{e_{-\chi}(k_{2}-k_{1})}{[4\kappa_+e_\chi+
4\kappa_-e_{-\chi}+K^2]^{3/2}},\\
\\
F_{k_{-1},k_{2}}({\bf k})=
-\frac{e_{-\chi}(e^{\chi}k_1+k_{-2})}{[4\kappa_+e_\chi+
4\kappa_-e_{-\chi}+K^2]^{3/2}}.
\end{array}
\label{bercur3}
\end{equation}

In conclusion, we demonstrated that, by analogy to the SU(2) group, the anholonomy of the SL(2,R) evolution can also be illustrated as a rotation of a parallel transported triad along a curve, but in the 3D Minkovsky space.
Several theoretical results for the SU(2) group have been derived using such an interpretation \cite{dandoloff-92jpa}, and one can attempt to derive similar expressions for the non-Hermitian 
evolution, however, we do not pursue them here.
Instead, we pointed out that the
model of a stochastic pump, developed in Ref.~\onlinecite{sinitsyn-07epl} leads to an evolution described by the SL(2,R) group, and we used it to 
derive all components of the Berry curvature in the parameter space. Our work should help further understanding of the stochastic pump effect.
For example, the non-adiabatic extension of the SL(2,R) geometric phase has been studied previously \cite{gao}. It should be possible to transfer some of the results of that study to
the problem of driven stochastic transport, and thus extend the recent progress on stochastic pump effect in the non-adiabatic regime \cite{ohkubo-08}.
 It would also  be important to find out whether the quantization of stochastic pump currents \cite{astumian-quantized} can be related to topological properties of the underlying symmetry group of evolution of the moments generating
function.  

\begin{acknowledgments} 
{\it  This work was funded in part by DOE under Contract No.
  DE-AC52-06NA25396.} 
\end{acknowledgments}


\begin{thebibliography}{99}
\bibitem{comment} In Ref.~\cite{sinitsyn-07epl} the parameter $\chi$ is defined to be multiplied by an imaginary unit $\chi \rightarrow i\chi$. It is, however, clear that making it imaginary played purely
decorative role in the discussion, so we switch to the description of the model using only real parameters in order to highlight the relation to SL(2,R) group.
\end{thebibliography}
\end{document}